\documentstyle[preprint,prl,aps,epsfig]{revtex}
\begin{document}
\title {Non-Abelian phases, charge pumping, and quantum computation with
Josephson junctions}
\author {Lara Faoro$^{(1)}$, Jens Siewert$^{(2,4)}$,
        and Rosario Fazio$^{(3,4)}$}
\address{$^{(1)}$ INFM $\&$ Institute for Scientific Interchange (ISI),
Viale Settimo Severo 65, 10133 Torino, Italy \\
$^{(2)}$ Institut f\"ur Theoretische Physik, Universit\"at Regensburg,
93040 Regensburg, Germany\\
$^{(3)}$ NEST-INFM $\&$ Scuola Normale Superiore,
56126 Pisa, Italy\thanks{permanent address}\\
$^{(4)}$ Dipartimento di Metodologie Fisiche e Chimiche (DMFCI),
        Universit\`a di Catania, viale A. Doria 6, 95125 Catania, Italy}

\maketitle

\begin{abstract}
Non-Abelian geometric phases can be generated and detected
in certain superconducting nanocircuits. Here we consider
an example where the holonomies are related to the adiabatic 
charge dynamics of the Josephson network.
We demonstrate that such a device can be applied both
for adiabatic charge pumping and as an implementation 
of a quantum computer.
\end{abstract}

\newpage

If a quantum system is prepared in a superposition of two states,
a physical observable associated with this system can exhibit oscillatory
behavior depending on the relative phase of the two states.
Interference can be induced during the dynamical evolution of the system, 
in this case we refer to the accumulated phase as the {\em dynamical} phase.
Interference can also be of {\em geometrical} nature if the
parameters (coupling constants, external fields,$\dots$) of the Hamiltonian are 
varied cyclically~\cite{berry84}. 
After Berry's original work, considerable attention has been devoted to the
interpretation, generalization, and detection of geometric 
phases~\cite{shapere89}. An important generalization is when
the adiabatic cyclic evolution involves a degenerate eigenspace of the
Hamiltonian. In this case it has been shown by Wilczek and Zee~\cite{wilczek84} 
that the evolution over a closed path does not result in a phase 
change but it leads to a superposition of the degenerate eigenstates
and the geometric phase acquires a non-Abelian structure.
Originally investigated in Nuclear Quadrupole Resonance~\cite{tycko87}, 
more recently it was shown that non-Abelian phases occur in the manipulation 
of trapped ions~\cite{unanyan99,duan01}.

Apart from its fundamental importance, geometric interference has interesting 
applications in the field of quantum information processing~\cite{nielsen00,ekert96}.
Implementations of quantum computers so far include
optical systems and liquid-state NMR~\cite{divincenzo00} as well as
solid-state devices based on superconductors~\cite{makhlin01} and on
semiconductors~\cite{loss00}. 
Recently it has been shown that quantum computation can also be 
implemented by geometric means ({\em geometric quantum computation}) 
using Abelian~\cite{abelian}
as well as non-Abelian~\cite{zanardi99,duan01} phases. 

Non-Abelian phases can also appear in the quantum dynamics of superconducting 
nanocircuits~\cite{choi01}, this is what we will show in this work.
There are various interesting aspects associated with this analysis. 
In addition to their possible detection, which is intriguing by itself, 
the existence of non-Abelian phases in superconducting nanocircuits leads to 
a new scheme for adiabatic charge pumping and allows to implement 
solid state holonomic quantum computation. Some parts of our proposal are on 
purpose speculative. The adiabatic manipulation of degenerate subspaces 
and the degeneracy condition itself is non-trivial to achieve for an 
artificially fabricated device.
We believe, however, due to the rapid development in the 
control of artificial two-level systems in solid-state
devices~\cite{solidtls}, the realization of geometric 
interference in mesoscopic systems has become plausible. 
Furthermore the applications of non-Abelian phases in pumping and 
computation are interesting new directions to pursue.

\begin{figure}
\centerline{
\epsfig{figure=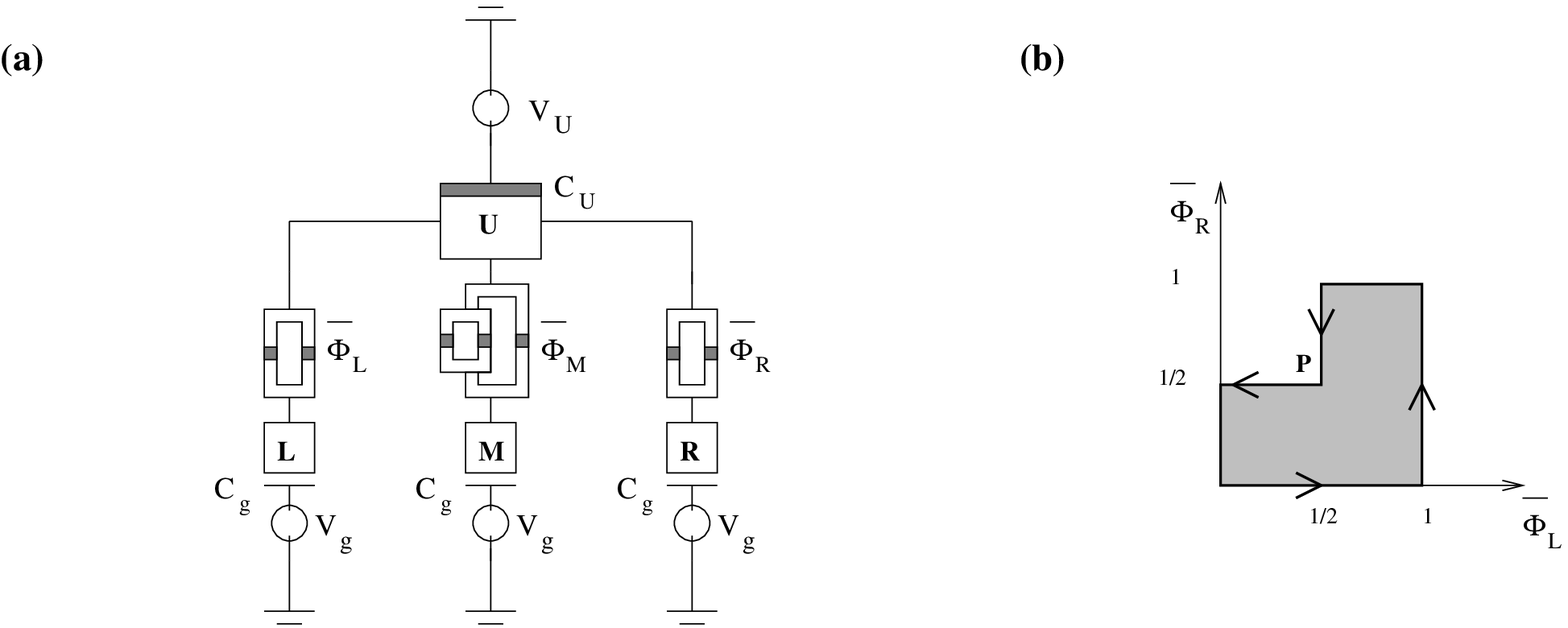, width=1 \textwidth}}
\vspace*{1.0cm}
\caption{a)Elementary Josephson network for the investigation of
         non-Abelian geometric phases.
	Note that an asymmetric SQUID loop cannot be switched off
	completely~\protect\cite{footnote1}.
	Since $J_{\rm M}=0$ may be desirable for quantum computation
	the SQUID is designed such that this condition can be satisfied;
	b)Pumping cycle for $3$ islands.
	Starting from $P=(1/2,1/2)$ and adiabatically following the drawn path,
	the gate $\displaystyle{U=e^{i \frac{\pi}{2} \sigma_{y}}}$ can be 
	achieved.}
\label{setup}
\end{figure}

In our discussion of non-Abelian phases in Josephson junction
circuits we follow the spirit of the
schemes described in Refs.\cite{unanyan99,duan01}. 
The starting point is the network shown in Fig.\ref{setup}a).
It consists of three superconducting
islands labeled by $j=L,M,R$ (Left, Middle, Right)
each of which is connected to a fourth (Upper) island labeled with $U$.
Gate voltages are applied to the three bottom islands via gate 
capacitances. 
The device
operates in the charging regime, that is the Josephson energies
$J_j$ ($j=L,M,R$) of the
junctions are much smaller than the charging energy $E_C$ of the
setup.
Each coupling is designed 
as a Josephson interferometer (a loop interrupted by two 
junctions and pierced by a magnetic field) as shown in Fig.\ref{setup}a.
Thus the effective Josephson energies $J_j$ can be tuned by changing 
the flux in the corresponding loop.
Electrostatic energies can be varied by changing the gate voltages 
$V_{g}$. 

Let us first analyze the electrostatic problem (i.e.\ $J_j\equiv 0$). 
For the sake of simplicity we assume that all capacitances are equal 
to $C$ and we consider identical gate charges $\displaystyle{q_g=C_g V_g/(2e)}$
for the three bottom islands. The charge states are indicated as
$\mid n_U,n_{L},n_{M},n_{R}\rangle$ where $n$ labels the 
number of Cooper pairs in the corresponding island~\cite{footnote}.
For gate charges $\displaystyle{ q_g \simeq 1/2}$ 
and $\displaystyle{1<2q_U+3q_g<2}$ (where
$\displaystyle{q_U=C_U V_U/(2e)}$, see Fig.\ref{setup}a), only four charge 
states are important as long as $T \ll E_C=(2e)^2/(4C)$. 
Three of these charge states, 
$\mid 0,1,0,0\rangle,\mid 0,0,1,0\rangle,\mid 0,0,0,1\rangle$ 
are degenerate. Their charge configuration corresponds to
one excess Cooper pair in one of the islands $j=L,C,R$, 
and none in the island $U$. 
The fourth state $\mid 1,0,0,0\rangle$ has one excess pair
on the island $U$ and none on the other islands.
All other charge states are much higher in energy.

The Josephson couplings $J_j$ allow for tunneling between the upper 
island and each of the bottom islands. 
The quantum-mechanical Hamiltonian of this simple four-state system reads
(in complete analogy with Refs.\cite{unanyan99,duan01})\pagebreak
\begin{eqnarray}
&&H=\delta E_{C}|1,0,0,0\rangle \langle 1,0,0,0| + \frac{1}{2} \bigl [
J_L(\bar{\Phi}_L) |1,0,0,0\rangle \langle 0,1,0,0|\ + \, \nonumber \\
&&~~~~+ J_{M}(\bar{\Phi}_M) |1,0,0,0\rangle \langle 0,0,1,0|
+ J_R(\bar{\Phi}_R) |1,0,0,0\rangle \langle 0,0,0,1|
+ {\rm H.c.\/} \bigr ]\,
\label{ham}
\end{eqnarray}
where
$\displaystyle{\delta E_{C}= \frac{4}{5} E_C\left [(q_g-q_U)
-\frac{1}{2}\right ]}$ is the energy difference between the 
three degenerate states and the fourth one, 
and $\bar{\Phi}_j=\Phi_j/\Phi_0$ are the external magnetic 
fluxes in units of the flux quantum $\Phi_0=hc/2e$~\cite{footnote1}.
In all the manipulations described below the gate voltages will be kept 
fixed. The three fluxes $\displaystyle{\{ \bar{\Phi}_L,\bar{\Phi}_M, 
\bar{\Phi}_R \}}$ are the parameters which will be varied cyclically. 
In general all the SQUID loops could be asymmetric, although it is
not necessary for the purpose of our discussion.

The Hamiltonian defined in Eq.(\ref{ham}) can easily be diagonalized. 
The lowest and highest eigenstate are non-degenerate.
The peculiar feature, 
exploited in \cite{unanyan99,duan01} is that 
the other two states (with zero energy) are degenerate for arbitrary
values of the couplings $J_j$. The subspace is 
spanned by the eigenstates (not normalized)
\begin{eqnarray}
&&|D_1\rangle=-J_{M} |0,1,0,0\rangle\ + J_L
 |0,0,1,0\rangle \ , \nonumber\\
&&|D_2\rangle = -J_R \left (
 J_L^* |0,1,0,0\rangle + J_{M}^*
  |0,0,1,0\rangle\right ) + (|J_L|^2+|J_M|^2) |0,0,0,1\rangle.
\label{aut}
\end{eqnarray}
By manipulation of the external magnetic fluxes it is possible
to generate non-Abelian phases.
We will show that by means of such phases adiabatic charge pumping and 
holonomic quantum computation can be realized 
with superconducting nanocircuits.

\underline{Charge pumping} -
For this purpose it is sufficient to have only symmetric SQUID loops.
In contrast to the well-known turnstiles for single electrons
or Cooper pairs~\cite{pothier91,geerligs91,pekola99} 
(in which the gate potentials are modulated periodically), here charge is
transported through the chain (from the L-island to the R-island)
by means of modulating the Josephson couplings while keeping the
gate voltages unchanged. The pumping cycle goes as follows.
The system is initially prepared in the $|0,1,0,0\rangle$ state
(i.e.the state $|D_1\rangle$ with $J_L ,J_M =0$ ) where the 
Cooper pair is in the left island. This can be achieved by 
turning off all Josephson couplings and coupling the L-island to a 
lead which provides the extra Cooper pair. Once the charge is on the
island,
one should change adiabatically the magnetic fluxes along a 
closed loop $\gamma$. At the end of the loop the initially prepared stated 
$|D_1\rangle$ will be mapped into the following rotated state:
\(
|D_1\rangle \longrightarrow U_{\gamma} |D_1\rangle ,
\)
where the unitary matrix $U_\gamma$ may be expressed as~\cite{wilczek84}:
\begin{equation}
U_\gamma={\bf P} \exp \oint_{\gamma} \sum_{j=L,M,R} A_j d \bar{\Phi}_j \, ; ~~~
 (A_j)_{\alpha,\beta}=\langle D_\alpha| \frac{\partial}{\partial \bar{\Phi}_j}|D_\beta \rangle,~~~ 
 \alpha ,\beta =1,2. \;
 \label{hol} 
 \end{equation}
Path ordering {\bf P} is required as, in general, 
the matrices $A_j$ do not
commute along the path.
If the path $\gamma$ is chosen
in the $(\bar{\Phi}_L,\bar{\Phi}_R)$- plane (at fixed
$\bar{\Phi}_M=0$), as shown in Fig.\ref{setup}b it can be shown that, 
after one adiabatic cycle, the final 
state of the system is $|0,0,0,1\rangle$, i.e. one Cooper pair 
has been transported 
through the chain of the three islands~\cite{altshuler99}.

The mechanism described here relies entirely on the geometric phase
accumulated during the cycle and can be generalized to
describe pumping of a single Cooper pair through $N$ superconducting islands.
The connection between pumping and geometric phases has been
discussed by Pekola {\em et al.}~\cite{pekola99}.
The crucial difference is that here only the
Josephson couplings have to be varied. During the cycle, exactly one Cooper
pair is transported, in this sense there are no errors due to the spread
of the wave function discussed in ~\cite{pekola99}.
There are drawbacks though, mostly related to the fact 
that the degenerate states are not the ground state and relaxation 
processes may become important.

\underline{Quantum computation, one qubit} -
The pumping process illustrated so far is nothing but one of the key elements 
to construct a quantum computing scheme using non-Abelian phases.
Proceeding along the lines of Ref.\cite{duan01},
we point out the necessary ingredients and the differences which
arise in the case of the Josephson junction setup. The nanocircuit presented in 
Fig.\ref{setup}a constitutes the qubit. The logical states 
to encode information in this implementation are
\begin{eqnarray}
|0\rangle_{\ell} &=&|0,1,0,0\rangle \nonumber \\
|1\rangle_{\ell} &=&|0,0,0,1\rangle \nonumber \ \ .
\end{eqnarray}
The other two charge states ($|1,0,0,0\rangle$ and $|0,0,1,0\rangle$) 
serve as auxiliary states. 
To show that the implementation is possible, it is sufficient to provide explicit
representations for the gates $U_1=e^{i \Sigma_1 |1\rangle_{\ell\ell}\langle 1|}$ and
$U_2=e^{i \Sigma_2 \sigma_y}$, describing rotations of
the qubit state about the $z$ axis and the $y$ axis, respectively.
In this case only one asymmetric SQUID 
(as shown in Fig.\ref{setup}) is required to implement the 
one-qubit operations.

\begin{figure}
\vspace*{2.5cm}
\centerline{
\epsfig{figure=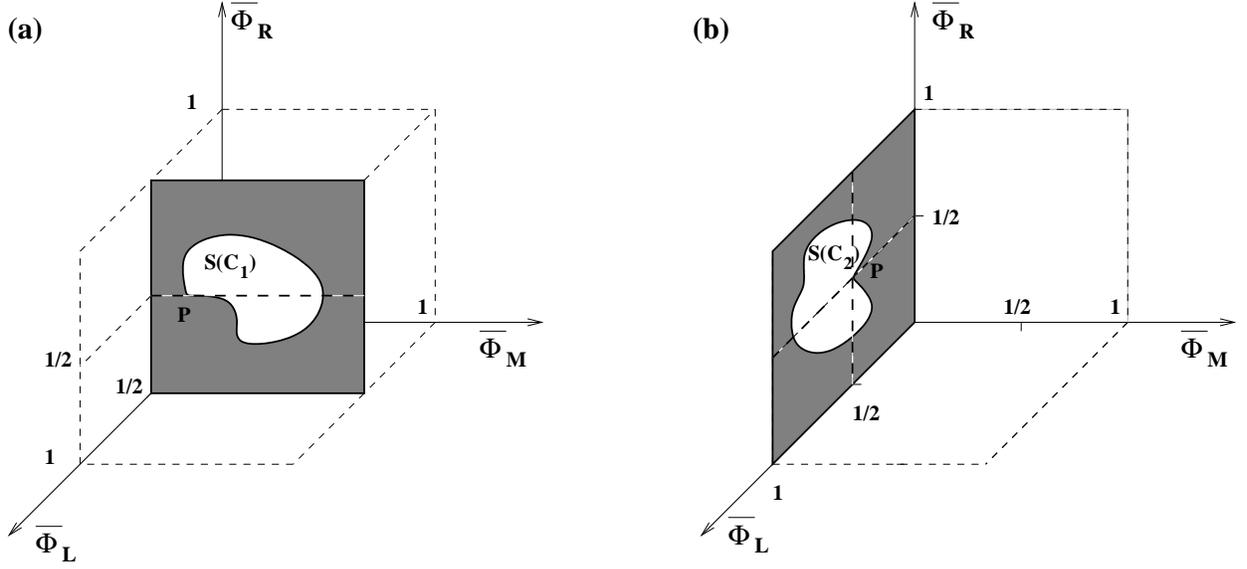, width=1 \textwidth}}
\vspace*{1.0cm}
\caption{
	Geometric realizations of gates $U_1=e^{i \Sigma_1 
	|1\rangle_{\ell\ell}\langle 1|}$ (a) and $U_2=e^{i
	\Sigma_2 \sigma_y}$ (b). The structure (non-zero elements) 
	of the unitary matrices $U_1$ and $U_2$ is 
	determined by the choice of the plane containing the loop and by the 
	starting/ending point of the 
	closed path. Different values of phase $\Sigma_1$ $(\Sigma_2)$ can be 
	obtained by varying the area 
	enclosed by loops $C_1$ $(C_2)$.}
\label{operations}
\end{figure}

The gate $U_1$ is a phase shift for the state $|1\rangle_{\ell}$
while the state $|0\rangle_{\ell}$ remains decoupled, i.e., $J_L\equiv0$ during
the operation.
In the initial state we have $J_R=0$,
so the eigenstates $\{|D_1\rangle, |D_2\rangle\}$ correspond to
the logical states $\{|0\rangle_{\ell}, |1\rangle_{\ell}\}$.
The control parameters $\bar{\Phi}_{M},\bar{\Phi}_R$
evolve adiabatically along the closed loop $C_1$ in the
$(\bar{\Phi}_{M}, \bar{\Phi}_R )$-plane from $\displaystyle{\bar{\Phi}_R=1/2}$
to $\displaystyle{\bar{\Phi}_R=1/2}$ (see Fig.\ref{operations}a).
By using the formula for holonomies Eq.(\ref{hol}) one can show that this
cyclic evolution produces the gate $U_1$ with the phase $\Sigma_1$ :
\begin{equation}
\Sigma_1 = \sigma_1
\oint_{{\cal S}(C_1)} d
\bar{\Phi}_{\rm M} d\bar{\Phi}_{R}
\frac{\sin \left (2 \pi \bar{\Phi}_R \right)}
{\left (|J_R(\bar{\Phi}_R)|^2+ |J_M(\bar{\Phi}_M)|^2\right )^2}\,
\nonumber
\end{equation}
where ${\cal S}(C_1)$ denotes the surface enclosed by the loop
$C_1$ in ${\cal M}$. 
and $\sigma_1=4 \pi^2 J_R(0)^2 (|J_{Ml}|^2-|J_{Mr}|^2)$. 
\\
Similarly we can consider a closed loop
$C_2$ (see Fig.\ref{operations}b) in the $(\bar{\Phi}_L,\bar{\Phi}_R)$-plane at fixed
$\bar{\Phi}_{\rm M}=0$,
and let the control parameters $\bar{\Phi}_L$ and $\bar{\Phi}_R$ undergo
a cyclic adiabatic evolution with starting and ending point
$\displaystyle{\bar{\Phi}_L=\bar{\Phi}_R=1/2}$.
This operation yields the gate $U_2$ with phase $\Sigma_2$
\begin{equation}
\Sigma_2 = \sigma_2
\oint_{{\cal S}(C_2)} d\bar{\Phi}_R d\bar{\Phi}_L
\frac{\sin \left(\pi \bar{\Phi}_L\right) \sin \left(\pi \bar{\Phi}_R
\right)} {\left ( J_M(0)^2 + |J_R(\bar{\Phi}_R)|^2 + |J_L(\bar{\Phi}_L)|^2
\right)^{3/2}}
\end{equation}
where ${\cal S}(C_2)$ denotes the surface enclosed by the loop
$C_2$ in ${\cal M}$, 
and $\sigma_2 = 4 \pi^2 |J_R(0)|^2 (|J_{Ml}|^2+|J_{Mr}|^2)$ where
we have assumed $J_L(0)=J_R(0)$. 
Obviously the pumping
cycle discussed above is a special case of the gate $U_2$ with
$\Sigma_2=\pi/2$.

\underline{Quantum computation, two qubits} -
It turns out that it is possible to implement a
conditional phase shift $U_3=e^{i\Sigma_3 |11\rangle_{\ell\ell}\langle11|}$
by coupling two qubits via Josephson junctions.
These junctions should be realized as symmetric SQUID loops such
that the coupling can be switched off.
The capacitive coupling due to these SQUID loops
can be neglected if the capacitances of the junctions are
sufficiently small~\cite{siewert01}.

\begin{figure}
\vspace*{2.5cm}
\centerline{
\epsfig{figure=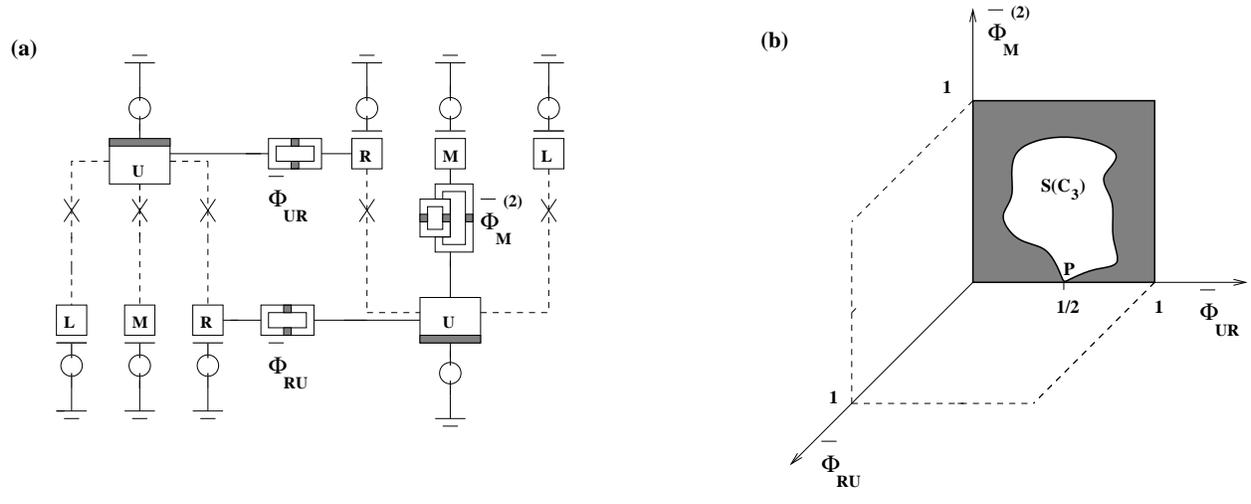, width=1 \textwidth}}
\vspace*{0.5cm}
\caption{Inter-qubit coupling for the implementation of the gate $U_3$ 
	and its geometric realization.}
\label{two-bit}
\end{figure}

By setting $\delta E_C=0$ (this was not
necessary in the one-qubit case)
and by coupling the qubits as shown in Fig.\ref{two-bit}a) 
we obtain the Hamiltonian
\begin{equation}
H_{\rm 2qubit} = \frac{1}{2}\left[
                     J^{(2)}_{M}
                     |{UU}\rangle \langle {UM}|
             + J_X |11\rangle_{\ell}\langle {UU}|+ {\rm H.c.} \right]\
\label{2bithamilt}
\end{equation}
where we introduced the notation $|U\rangle =|1,0,0,0\rangle$ 
and $|M\rangle =|0,0,1,0\rangle$ for the auxiliary
states which are coupled by the interaction between the qubits.
The matrix element $J_X=J_X(\bar{\Phi}_{\em UR},\bar{\Phi}_{\em RU})$ is given by
$J_X =-(1/2) J_{UR}(\bar{\Phi}_{\em UR})\, J_{\em RU}(\bar{\Phi}_{\em UR})\,\mu$
where $\displaystyle{\mu=\left[1/\delta E_C^{+-}+
                              1/\delta E_C^{-+}\right]}$.
Here $\delta E_C^{+-}$ and $\delta E_C^{-+}$ denote the 
charging energy difference between the initial and the intermediate
state (see below).
The coupling is of second order in the Josephson energies since the
inter-qubit coupling junctions change the total number
of pairs on each one-bit setup. Thus the coupling occurs
via intermediate charge states which lie outside the Hilbert space
of the two-qubit system. These are states, 
e.g.\ $|0,0,0,0\rangle \otimes |1,0,0,1\rangle$,
without excess Cooper pair on
the first qubit and two excess pairs on the second qubit. 
We have abbreviated the charging energy difference
between the corresponding state and the initial qubit state by
$\delta E_C^{-+}$, and we have denoted the external magnetic
fluxes in the coupling SQUID loops by $\Phi_{\em UR}$ and $\Phi_{\em RU}$.\\
\\
While $J_X(\bar{\Phi}_{\em UR},\bar{\Phi}_{\em RU})$ is the only off-diagonal
coupling of second order, there are also second-order corrections of
the diagonal elements, i.e.\ of the energies of the two-qubit states.
These corrections would lift the degeneracy and thus would hamper
the geometric operation which is based on the degeneracy of all states.
It is therefore crucial that it is possible to compensate these corrections
and to guarantee the degeneracy. It is easy to see that by adjusting
the gate voltages the energy shifts can be canceled.
Note that during the geometric operation the values of the Josephson
couplings are changing and therefore also the energy shifts are
not constant. Consequently their compensation by means of the gate
voltages has to follow the evolution of the parameters.
\\
Let us now show explicitly how the gate $U_3$ can be achieved.
To this aim, we consider a closed loop $C_3$ in the
$(\bar{\Phi}^{(2)}_{\em M}, \bar{\Phi}_{\em UR})$-plane at
fixed $\bar{\Phi}_{\em RU}=0$. (See Fig.\ref{two-bit}b). If the control parameters
$\bar{\Phi}_{\em UR}$ and $\bar{\Phi}_{\em M}^{(2)}$ undergo a cyclic adiabatic
evolution with starting and ending point
$\bar{\Phi}_{\em UR}=1/2$, $\bar{\Phi}_{\em M}^{(2)}=0$, the
geometric phase obtained with this loop is
$$
\Sigma_3 = \sigma_3
\oint_{{\cal S}(C_3)} d\bar{\Phi}_{ M}^{(2)} d\bar{\Phi}_{ UR}
\frac{ \mu^2 \sin \left(2 \pi \bar{\Phi}_{UR}\right)} {\left (
|J_{M}^{(2)}(\bar{\Phi}_{M}^{(2)})|^2+
\mu^2 \, J_{RU}(0)^2 |J_{UR}(\bar{\Phi}_{UR})|\right)^2}
$$
with $\sigma_3 = 4 \pi^2 J_{RU}(0)^4 (|J_{{M}l|}^2-|J_{{M}r}|^2)$ and $J_{ UR}(0)=
J_{RU}(0)$.

As we have mentioned in the introduction, some caution is required
before regarding this scheme ready for implementation.
In practice it will be difficult to achieve perfect degeneracy
of all states. Thus the question is imposed to which extend incomplete
degeneracy of the qubit states is permissible. Clearly, the adiabatic
condition requires the inverse operation time $\tau_{\rm op}$
to be smaller than the minimum energy difference to the neighboring states:
$
 \tau_{\rm op}^{-1} \ll \min{\delta E_C,J_j,J_X}
$.
On the other hand, if the degeneracy is not complete and the deviation
is of the order $\epsilon$ one can show by modifying the derivation
of Eq.\ (4) in Ref.\cite{zanardi99} that for
$
  \epsilon\ll\tau_{\rm op}^{-1}
$
the holonomies can be realized to a sufficient accuracy.
This inequality expresses the requirement that
the operation time be still small enough in order to not resolve
small level spacings of the order $\epsilon$.

There is another important constraint on $\tau_{\rm op}$. As
the degenerate states in Eq.\ (\ref{aut}) are different from
the ground state of the system, $\tau_{\rm op}$ must not be too
large in order to prevent inelastic relaxation. The main origin for
such relaxation processes is the coupling to a
low-impedance electromagnetic environment. We can estimate
the relaxation rate by $\Gamma_{\rm in} \sim E (R_{\rm env}/R_K)$
where $R_K=h/e^2$ is the quantum resistance and $E$ is on the order of
the Josephson energies $E\sim J_j, J_X$.
Thus it is not difficult
to satisfy the condition $\tau_{\rm op} < \Gamma_{\rm in}^{-1}$
experimentally. In fact, it has been found recently that inelastic
relaxation times in charge qubits can be made quite large
and exceed by far the typical dephasing times due to background charge 
fluctuations~\cite{nakamura01,paladino01,whitney01}. 

Both charge pumping and the implementation of quantum computing are related 
to coherent manipulations of charge states. Therefore as a readout one can 
use the scheme developed~\cite{makhlin01} to measure charge qubits. 
No additional difficulty is forecasted at this level.

Stimulating discussions with G.\ Falci, G.M.\ Palma, E.\ Paladino,
M.\ Rasetti and P. Zanardi are gratefully acknowledged.
We wish to thank M.-S.\ Choi for informing us
of his results. This work has been supported by
the EU (IST-FET SQUBIT) and by INFM-PRA-SSQI.


\begin{references}
\bibitem{footnote}
	We ignore quasiparticles since they are suppressed at low
	temperature.
\end{references}
\end{document}